\documentclass[]{spie}  

\usepackage{xcolor}
\usepackage{amsmath,amsfonts,amssymb}
\usepackage{graphicx}
\usepackage[colorlinks=true, allcolors=blue]{hyperref}
\usepackage{subfigure}
\usepackage[euler]{textgreek}

\title{Daytime calibration and testing of the Keck All sky Precision Adaptive Optics Tomography System}

\author[a]{Avinash Surendran}
\author[a]{Jacques R. Delorme}
\author[a, b]{Carlos M. Correia}
\author[a]{Steve Doyle}
\author[a]{Sam Ragland}
\author[a]{Paul Richards}
\author[a]{Peter Wizinowich}
\author[d]{Philip M. Hinz}
\author[d]{Daren Dillon}
\author[d]{Cesar Laguna}
\author[a, e]{Sylvain Cetre}
\author[a]{Scott Lilley}
\author[a]{Ed Wetherell}
\author[a]{Jason C. Y. Chin}
\author[a]{Eduardo Marin}

\affil[a]{WM Keck Observatory, Hawaii, USA}
\affil[b]{Space ODT, Portugal}
\affil[d]{Univ. of California, Santa Cruz, USA}
\affil[e]{Durham University, UK}

\authorinfo{Further author information: (Send correspondence to A. Surendran.)\\E-mail: asurendran@keck.hawaii.edu, asurendran89@gmail.com}

\begin{document} 
\maketitle

\begin{abstract}
The development of the Keck All sky Precision Adaptive optics (KAPA) project was initiated in September 2018 to upgrade the Keck I adaptive optics (AO) system to enable laser tomography adaptive optics (LTAO) with a four laser guide star (LGS) asterism. The project includes the replacement of the existing LMCT laser with a Toptica laser, the implementation of a new real-time controller (RTC) and wavefront sensor optics and camera, and a new daytime calibration and test platform to provide the required infrastructure for laser tomography. The work presented here describes the new daytime calibration infrastructure to test the performance for the KAPA tomographic algorithms. This paper outlines the hardware infrastructure for daytime calibration and performance assessment of tomographic algorithms. This includes the implementation of an asterism simulator having fiber-coupled light sources simulating four Laser Guide Stars (LGS) and two Natural Guide Stars (NGS) at the AO bench focus, as well as the upgrade of the existing TelSim on the AO bench to simulate focal anisoplanatism and wind driven atmospheric turbulence. A phase screen, that can be adjusted in effective altitude, is used to simulate wind speeds up to 10 m/s for a duration of upto 3 s.
  
\end{abstract}


\section{Introduction}
\label{section:intro}
\begin{figure}[t]
\label{fig:fig1}
\includegraphics[width=8cm]{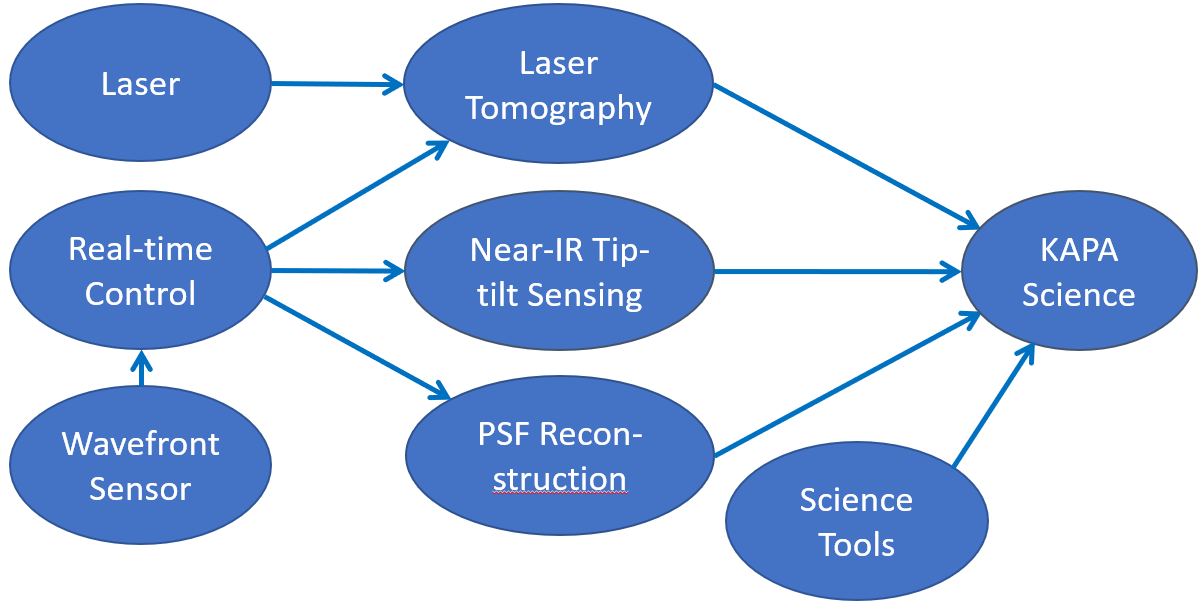}
\centering
\caption{KAPA Technical Elements}
\end{figure}

The Keck All sky Precision Adaptive optics (KAPA)\cite{wizinowich2020,wizinowich22} system is funded by the NSF Mid-Scale Innovation Program (MSIP) and was initiated in 2019. The KAPA system includes the following upgrades to the Keck 1 Adaptive Optics (AO) system (outlined in Figure \ref{fig:fig1}): 
\begin{itemize}
\item A new laser system including replacement of the existing laser with a laser from TOPTICA photonics (similar to the upgrade at Keck II\cite{chin2016}), and associated changes to the beam train and infrastructure. 
\item A real-time controller (RTC)\cite{Chin22} and wavefront sensor camera (First Light Optics OCAM2K\cite{ocam2k, philippe2014}) upgrade to provide the required base for laser tomography. This upgrade is a duplicate of an NSF MRI-funded upgrade on the Keck II telescope.
\item A laser tomography upgrade involving modifications to the new RTC for tomography, a system to generate an asterism of four LGS on a 7.6 arcsec radius, new wavefront sensor opto-mechanics to support four LGS on a single camera, and an asterism generator for daytime calibrations.
\item Point Spread Function reconstruction (PSF-R) for the OH-Suppressing Infra-Red Imaging Spectrograph (OSIRIS)\cite{larkin2006} science instrument. 
\end{itemize}

The main objectives of the KAPA daytime calibrations infrastructure are broadly divided as follows:
\begin{itemize}
\item Accurate calibrations for the NGS and single LGS AO modes by providing a diffraction-limited spectrally uniform output for the NGS source, and a seeing-limited 590nm output for the LGS source.
\item Use of a turbulence simulator with a translatable atmospheric phase screen for testing of tomographic reconstruction algorithms for laser tomography adaptive optics (LTAO) and error budget determination and troubleshooting for all AO modes.
\item Alignment and testing of the KAPA pupil relay optics.
\end{itemize}

\begin{figure}[t]
\includegraphics[width=6cm]{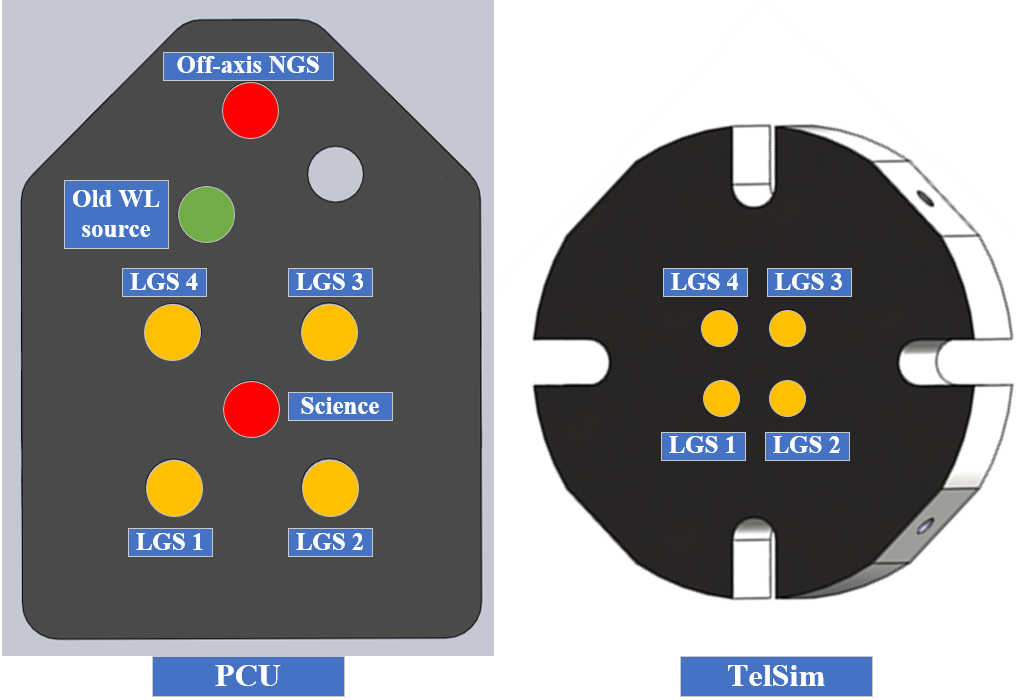}
\centering
\caption{Positions of the fibers outputs on the fiber holder and the light sources to which they are connected. The left image shows the fiber holder mounted on the PCU stage while the right image shows the asterism fiber holder mounted at the LGS focus of the TelSim.}
\label{fig:fig2}
\end{figure}

The preparation for LTAO daytime calibration and testing involves the integration of the following new hardware (described in detail in Section \ref{section:hardware}) into the AO bench:
\begin{itemize}
\item \textbf{Asterism Simulator (AS):} The KAPA LGS asterism, an on-axis and off-axis NGS are simulated by the AS for the purpose of performing daytime calibration (computing centroid offsets, verification of optical registration etc.). The AS consists of the light sources, their automated control mechanism, a photonic switcher to switch the light between the calibration output and the Telescope Simulator (TelSim) and the fiber holders which create the required asterism (shown with labeled outputs in Figure \ref{fig:fig2}). Both fiber holders support four multimode (MM) fibers at the equivalent on-sky asterism radius of 7.6". The Precision Calibration Unit (PCU) fiber holder will be positioned at the input focal plane of the AO system using the PCU stage (explained later here). A separate single-mode fiber will be located at the on-axis NGS focus at the input to the TelSim, while the PCU mounted holder includes one on-axis and two off-axis single-mode fibers. One off-axis single-mode fiber is from the existing white light source to support initial integration activities. The ~15" off-axis fiber will be used as an off-axis NGS for testing purposes. The single-mode fibers are illuminated with a white light source, and the multi-mode fibers with a narrow-band source close to the LGS wavelength of 589 nm. The AS infrastructure will be described in Section \ref{section:AS}.
\item \textbf{Telescope Simulator (TelSim):} The existing TelSim at Keck will be upgraded to incorporate the addition of four LGS sources at a conjugate distance of 100 km (to simulate a sodium layer altitude of 90km at a zenith angle of 30 deg) and a turbulence simulator having a translatable atmospheric phase screen equivalent to the size of $\sim$3.5 Keck pupils. The phase screen can translate for a maximum of 2.2s at a maximum equivalent wind speed of 10m/s, and can be positioned at equivalent elevations between 5km and 12km. The light sources and their control for the TelSim will be provided by the AS. This module is described in detail in Section \ref{section:TS}.
\end{itemize}

\begin{figure}
    \centering
    \subfigure[]{\includegraphics[width=0.48\textwidth]{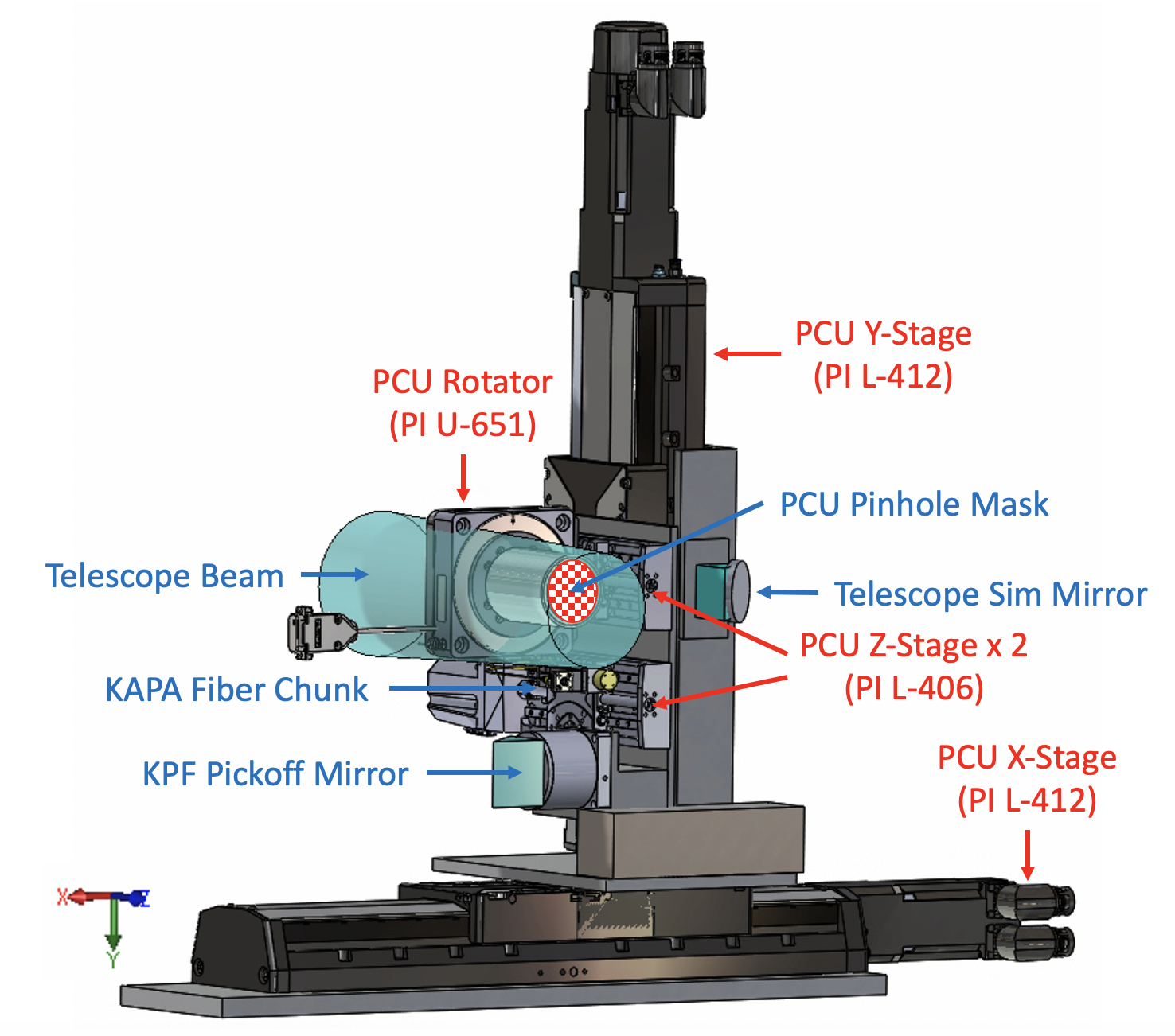}} 
    \subfigure[]{\includegraphics[width=0.43\textwidth]{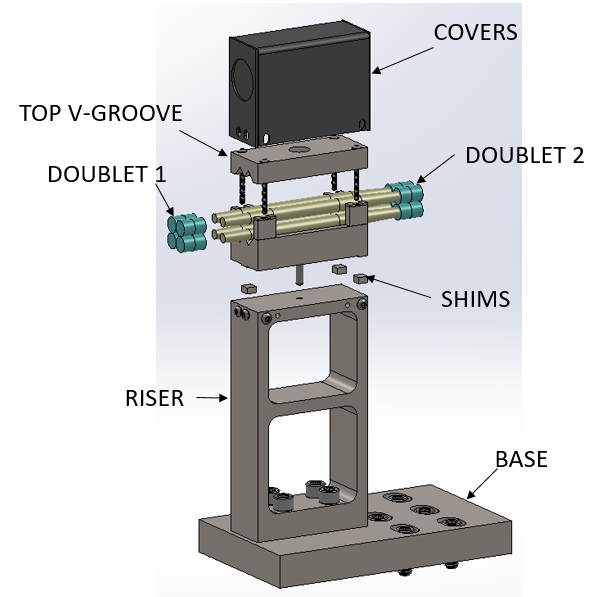}} 
    \caption{(a) The PCU, or Keck AO precision calibration unit is a project to replace the existing calibration unit and add support for new KAPA fibers, a pinhole mask, and fold mirrors for the TelSim and the fiber-injection unit (FIU) for the Keck Planet Finder (KPF) instrument (b) The KAPA Pupil Relay Optics (PRO) will produce four images of the Keck telescope pupil, one for each LGS, on the WFS lenslet array for WFS focus stage (FCS) positions conjugate to sources from 85 km to infinity.}
    \label{fig:PCU_PRO}
\end{figure}

The two other modifications in the AO bench, which are not described in this paper are the:
\begin{itemize}
\item \textbf{Precision Calibration Unit (PCU)\cite{Freeman21}:} The PCU will replace the existing Simulator Fiber Positioner (SFP). One AS fiber holder will be mounted on the PCU. The PCU will also support a fold mirror to the Keck Planet Finder Fiber Injection Unit and a pinhole mask in a rotation stage for OSIRIS distortion calibration. The schematic of the PCU is shown in Figure \ref{fig:PCU_PRO}a.
\item \textbf{New wavefront sensor with the new Pupil Relay Optics (PRO):} The wavefront sensor will be upgraded from the existing CCD39 to the FLI OCAM2k to accommodate the four LGS pupils on one camera and to provide better noise performance to compensate for the intensity reduction of the splitting of the single LGS into four, and to support the higher frame rate possible with the new RTC. The KAPA PRO will utilize a set of optics to relay the four LGS pupils to the lenslet array. This will be part of the array of optics in front of the new WFS to enable LT, which includes a new field stop and reducer optics in addition to the PRO. The schematic of the PRO is shown in Figure \ref{fig:PCU_PRO}b.
\end{itemize}

Section \ref{section:tomo} describes the changes in the AO high-order loop (as compared to the existing single NGS and LGS modes) that will enable tomography with the four LGS sources, and Section \ref{section:daytime} will describe the laser tomography algorithm validation and performance assessment tests to be performed with the daytime calibration infrastructure.

\section{Hardware}
\label{section:hardware}
\subsection{Asterism Simulator}
\label{section:AS}

\begin{figure}[t]
\includegraphics[width=0.8\textwidth]{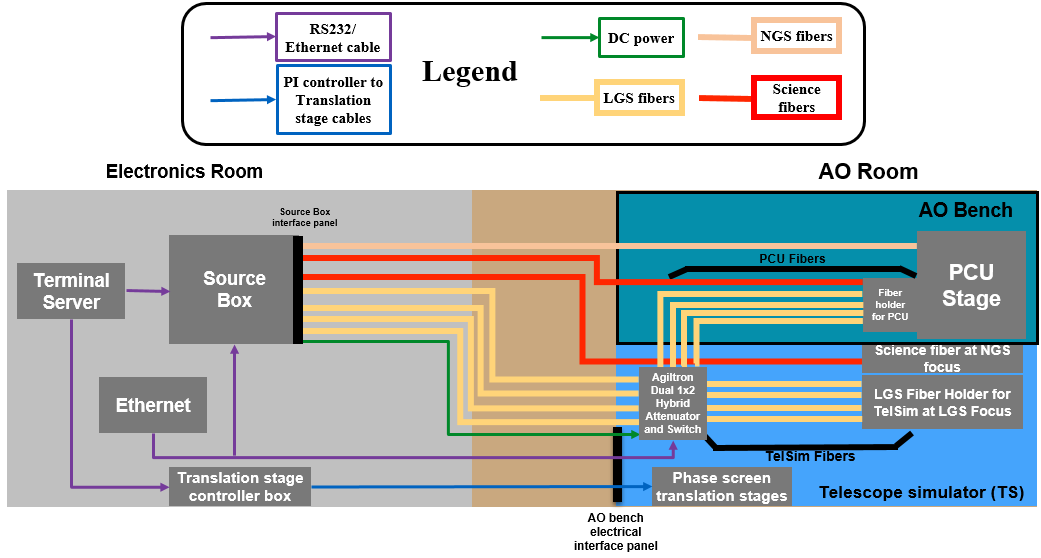}
\centering
\caption{Schematic of the KAPA Asterism Simulator. The source box (located in the electronics room) consists of all the light sources and their control hardware, with a fiber relay transporting the light to two locations (PCU and TelSim) on the AO bench.}
\label{fig:fig4}
\end{figure}

The function of the AS is to enable the following observation modes at the PCU output (used for normal daytime calibrations of the AO system) and at the TelSim output (used for LTAO performance assesment and error budget determination):
\begin{itemize}
\item LTAO Simulation Mode: 4x LGS, 1x Off-axis NGS, 1x Science. Switchable between the PCU and TelSim (without NGS) stages.
\item LGS AO Simulation Mode: 1x LGS, 1x Off-axis NGS, 1x Science. Switchable between the PCU and TelSim (without NGS) stages.
\item On-axis NGS AO Simulation Mode: 1x Science (acting as the on-axis NGS source). Switchable between the PCU and TelSim stages.
\item Off-axis NGS AO Simulation Mode: 1x NGS. Only available at the PCU stage.
\end{itemize}
As the AO bench is temperature controlled, all electronics are located in a nearby electronics vault 10 - 15 m away from the AO bench. The outputs of both the white light and the LGS sources (housed in a box inside the electronics vault) are sent to the PCU and TelSim fiber holders using fiber relays, as shown in Figure \ref{fig:fig4}. The white light sources (NGS and Science sources) should have a magnitude range of 9 in all wavelengths between the R-band and the K-band, while the LGS sources should have a magnitude range of 7 - 11 close to the sodium laser wavelength of 589 nm.

\subsubsection{Source Box}
\label{section:SB}
\begin{figure}
    \centering
    \subfigure[]{\includegraphics[width=0.38\textwidth]{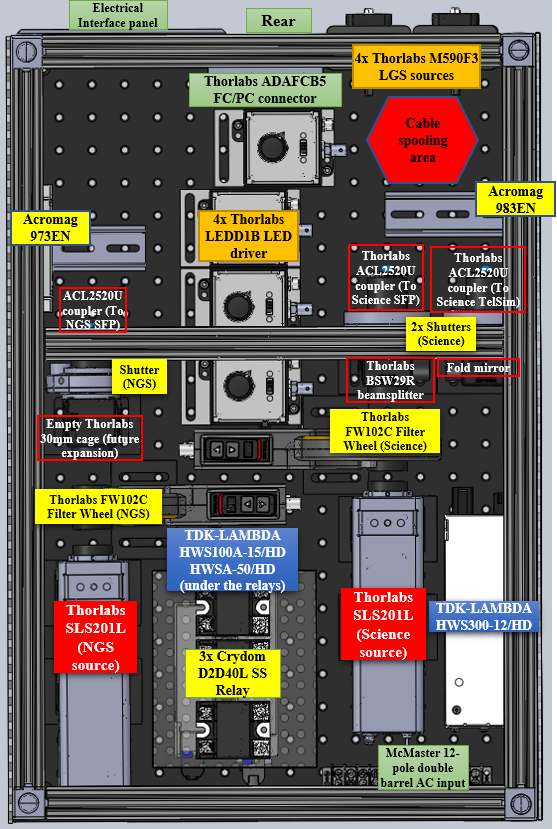}} 
    \subfigure[]{\includegraphics[width=0.58\textwidth]{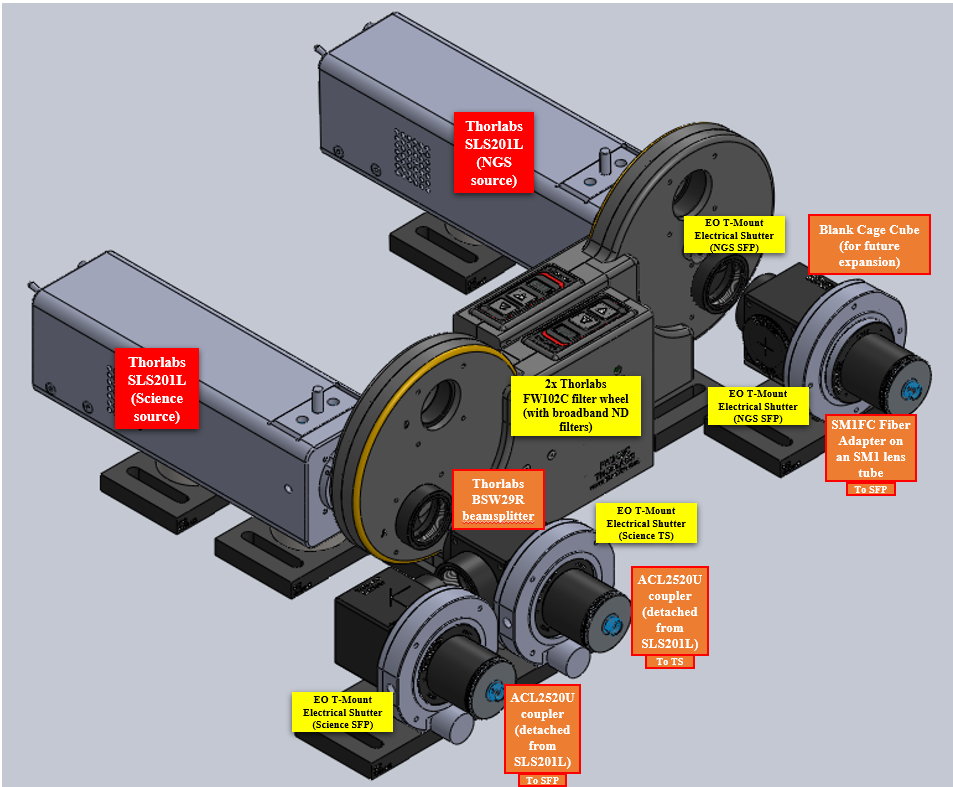}} 
    \caption{(a) Top view of solidworks layout of the source box with labelled components (b) Solidworks assembly and components for attenuation and switching for the NGS and Science sources}
    \label{fig:fig5}
\end{figure}

The source box consists of all the sources and their control mechanisms. The enclosure layout of the source box is shown in Figure \ref{fig:fig5}a and the white light arm of the source box is shown in Figure \ref{fig:fig5}b. The details of the sources and controllers are described below:
\begin{itemize}
    \item White light arm: The on-axis Science output and the off-axis NGS output are provided by two Thorlabs SLS201L tungsten-halogen sources. With a nominal wavelength range of 0.5 - 2.5 \textmu m, they provide enough spectral range for having sufficient light on the visible tip-tilt sensor (sensitive to the R-band) all the way through to the edge of the K-band of the science instrument, OSIRIS. A coverage of 9 magnitudes across the R-band to the K-band is provided by filter wheels with broadband neutral density (ND) filters, and a beamsplitter is used to split the Science output between the PCU and TelSim outputs. Shutters are used for the dual purposes of blocking light to one output while the other is being used, and are also used to minimize the number of times the sources are switched on and off, for ensuring the stability of output intensity and in extending the life of the source. Figure \ref{fig:fig5}b shows the source, the filter wheel, the beamsplitter and the shutters which are coupled to one of three outputs at the AO bench. An ethernet controlled digital PLC is used to control all shutters and solid state relays for turning on/off all the sources.
    \item LGS arm: The LGS outputs are provided by four Thorlabs M590F3 narrow-band (centered at 590 nm) fiber-coupled LED sources (acting as the LGS source). The LGS sources can be electronically controlled from zero to full intensity using a 0 - 5 V input voltage range, which is remotely provided through an ethernet controlled Analog PLC. A combined photonic attenuator and a 2-to-4 fiber splitter is used to split the four LGS sources among the PCU output and the TelSim output.
\end{itemize}

\begin{figure}[t]
\includegraphics[width=0.8\textwidth]{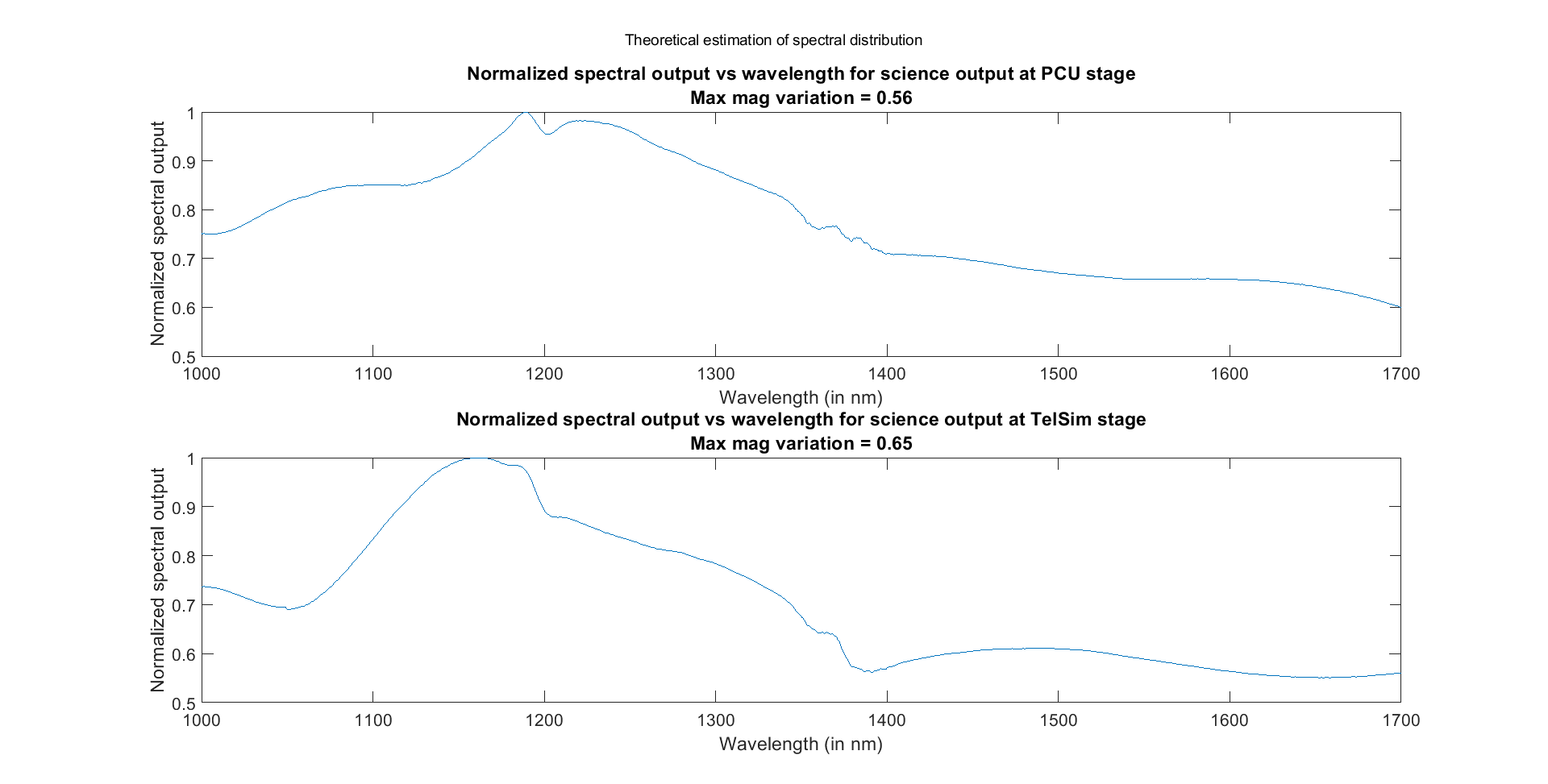}
\centering
\caption{Normalized spectral transmission for science source. Top figure shows the spectral variability estimate for the PCU stage output, and the bottom figure shows the same for the TelSim stage output.}
\label{fig:fig6}
\end{figure}

\subsubsection{Fiber Relay and Asterism Output}
\label{section:FR}
The fiber relay consists of the 12 - 15 m fibers used to connect the sources in the source box to the fiber holders in the AO bench. Four multi-mode fibers for the LGS and three single mode fibers for the white light sources form part of this relay. 

As with the white light arm of the source box, the primary design requirement for the white light fiber relay is to produce a diffraction limited output for wavelengths ranging from the R-band to the K-band with $<$ 2 magnitude variation between the wavelengths of 0.5 - 2.5 \textmu m. The requirements led to the use of using Zirconium Fluoride glass fibers that exhibit a high transparency from 0.3 to 4.5 µm. The most suitable fiber candidate that met the requirements was the ZFG SM [1.95] 6.5/125 fiber (manufactured by Le Verre Fluoré) owing to the consistent attenuation of $<$ 20dB in the wavelength range of 1.2 to 2.5 \textmu m and $<$ 30dB for 1.0 to 1.2 \textmu m, and a core and mode field diameter of 6.5 \textmu m and $<$ 10 \textmu m respectively (the diffraction limited spot requirement is $<$ 15 \textmu m). The combined spectral transmission of the white light source output, the beamsplitter, the ND filters and the ZFG fibers are estimated from their datasheets in Figure \ref{fig:fig6} and they satisfy the requirements between 1.0 to 1.7 \textmu m for which the data was available for all optical components.

The primary design requirement for the LGS fibers is to have a uniform beam illumination across the 800 \textmu m seeing-limited output at the AO bench focus and minimal attenuation at the LGS wavelength of 590 nm. The Thorlabs FT800UMT fibers were chosen for the LGS fiber relay.

\begin{figure}
    \centering
    \subfigure[]{\includegraphics[width=0.67\textwidth]{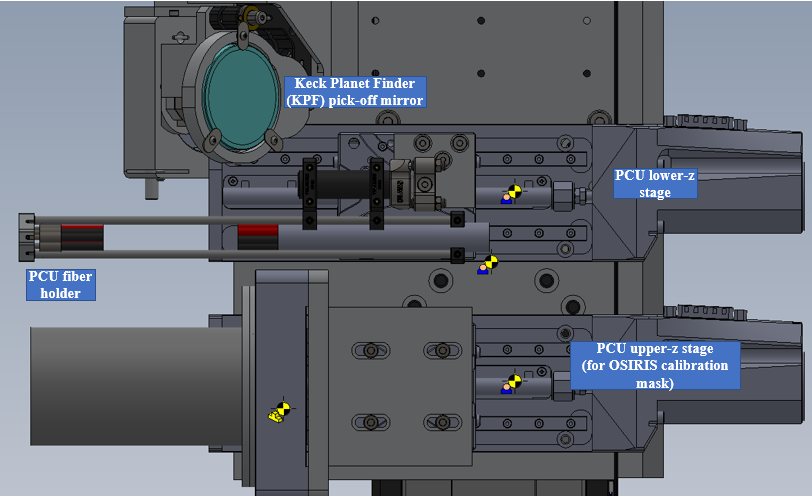}} 
    \subfigure[]{\includegraphics[width=0.25\textwidth]{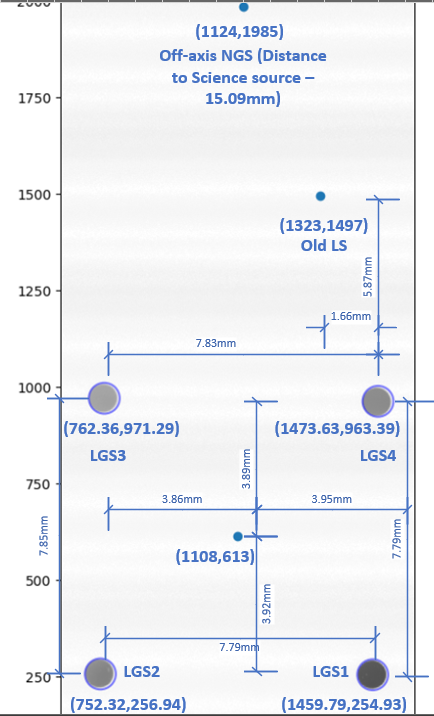}} 
    \caption{(a) Side view of the KAPA fiber holder mounted on the PCU lower-z stage (b) To measure the true fiber output separation, the fiber holder was imaged at a 1:1 magnification on an Andor Mirana sCMOS camera with a 11\textmu m pixel size, with the figure showing the centroid locations of each fiber output.}
    \label{fig:fig7}
\end{figure}

The schematics of the fiber holders on the PCU and the TelSim outputs are shown in Figure \ref{fig:fig2}. The fiber holders are machined to hold 2.5mm cylindrical ferrules at the equivalent asterism locations, and mounted on a standard 16mm cage system to be attached to the PCU lower z-stage (which is used to adjust the focus position of the fiber outputs). Figure \ref{fig:fig7}a shows the relative position of the PCU fiber holder with the surrounding components and Figure \ref{fig:fig7}b shows an image taken with an Andor camera after the PCU fiber outputs are positioned at the focus of a 1:1 optical relay. The target separation between the LGS fibers are 7.8 mm with the distance from the Science to the NGS source being 15 mm.

\subsection{Telescope and Turbulence Simulator}
\label{section:TS}

\subsubsection{Telescope Simulator}
\label{section:TelSim}

\begin{figure}[t]
\includegraphics[width=0.9\textwidth]{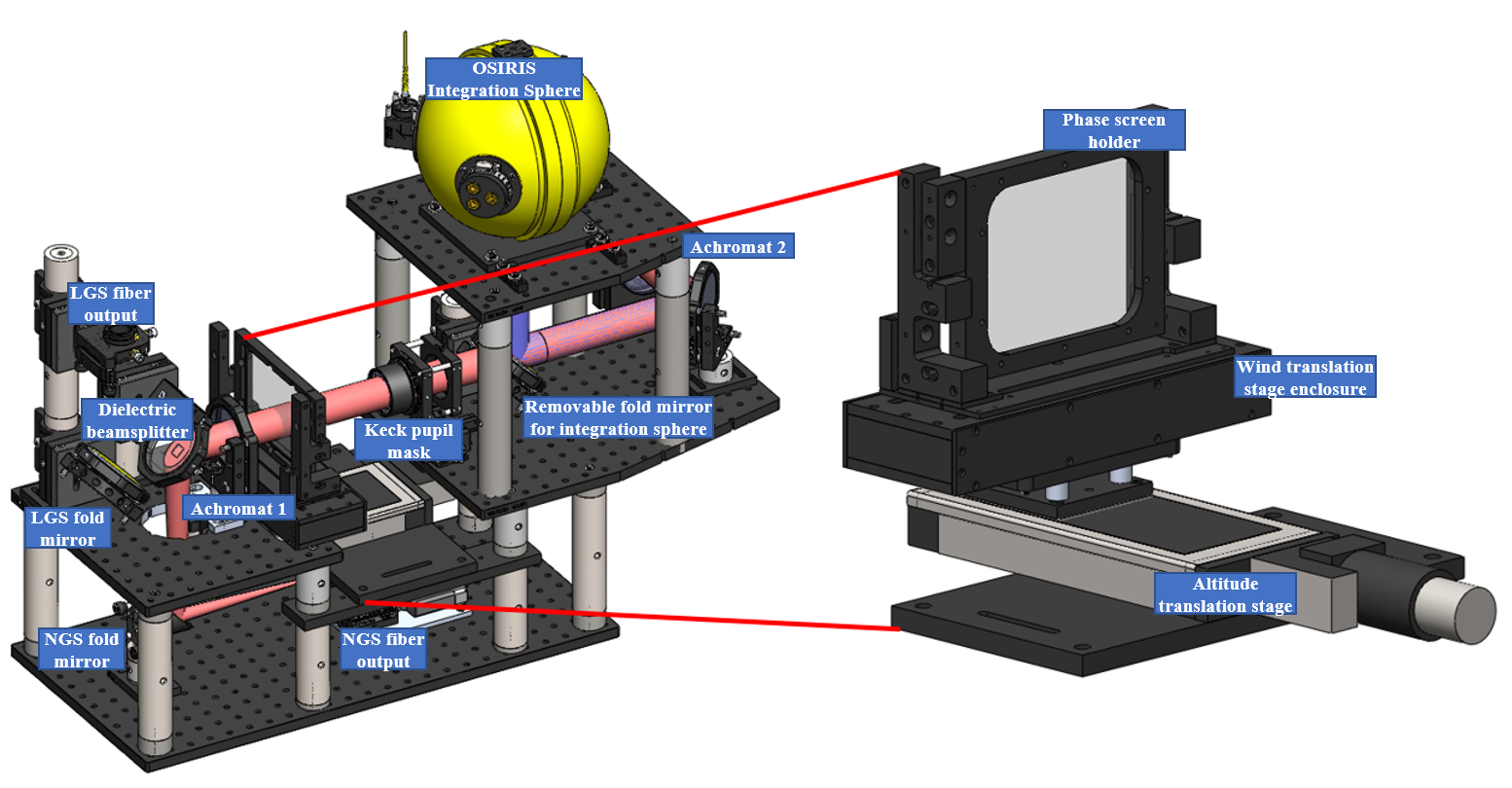}
\centering
\caption{CAD model of the TelSim (left) and the Turbulence Simulator (right)}
\label{fig:fig8}
\end{figure}

The existing TelSim was designed to implement a single fiber conjugate to the sodium layer, and a separate fiber at the NGS focus. The following changes have been made to the TelSim (Figure \ref{fig:fig8} left) for the purpose of validating the tomographic algorithms and in assessing the performance of the same:
\begin{itemize}
\item The TelSim optics consists of two identical achromats located two focal lengths apart with a pupil mask in the collimated space located between them. The pupil mask is conjugate to the same plane as the DM. 
\item The new design will consist of a fiber holder consisting of four fibers at the conjugate altitude of 100 km (90 km at a zenith angle of 30 degrees) and an on-axis science fiber at the NGS focus in the TelSim, with the two beams combined by a dielectric beamsplitter. The TelSim provides a Keck telescope pupil (as compared to the DM as the pupil stop for the PCU). 
\item While the DM will be able to simulate turbulence conjugate to the ground-layer, the turbulence which is conjugate to higher altitudes ($>$ 1 km) will be simulated by a phase screen adhering to the Kolmogorov spatial spectrum at the Mauna-Kea (MK) median seeing conditions. The phase screen will be positioned between the two achromats on the source side of the pupil mask, and it can be translated across different conjugate altitudes ranging from $\sim$ 5 - 15 km and across the pupil. This sub-assembly is the turbulence simulator (Figure \ref{fig:fig8} right).
\item Depending on whether the TelSim is being used for calibrating OSIRIS or for assessing the performance of the tomographic algorithms, a kinematic magnetic base is used to hold or remove a fold mirror to fold the beam coming from the integration sphere into the AO bench.
\end{itemize}

\subsubsection{Phase Screen}

\begin{figure}
    \centering
    \subfigure[]{\includegraphics[width=0.6\textwidth]{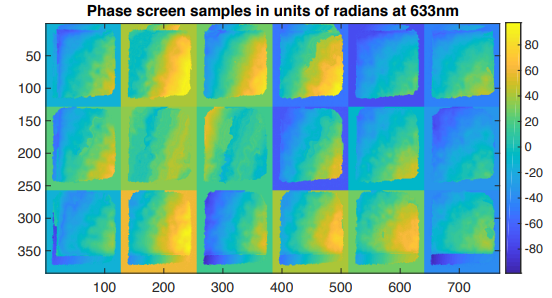}} 
    \subfigure[]{\includegraphics[width=0.48\textwidth]{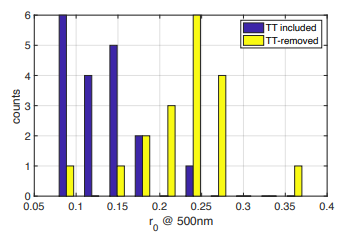}} 
    \subfigure[]{\includegraphics[width=0.48\textwidth]{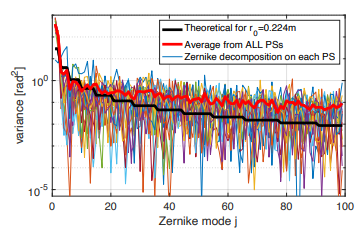}} 
    \caption{(a) QPI profile of the phase screen at a wavelength of 633nm, divided into 3.2 cm sections limited by the beam size (b) Histogram of $r_0$ estimated from fitting to Noll’s Zernike polynomial variances including tilt (blue) and tilt-removed (yellow) (c) Zernike modal decomposition and the average over all the turbulence snapshots. The theoretical curve for the prescribed turbulence is in black (scaled to the usable diameters cropped out of the empirical phase-screens).}
    \label{fig:fig9}
\end{figure}

The two options for the shape of the phase screen and the motion associated with wind velocity was either a circular phase screen with rotational motion or a rectangular phase screen with a linear motion respectively. Even with a maximum diameter of 150 mm for a circular phase screen (that most vendors were ready to fabricate with), the edge of the circular phase screen would have a relative speed of about two times the inner edge of the pupil, considering the pupil (having a diameter of 36.5mm) to be located at the edge of the circle. This would disable the option of having a controlled constant wind speed for LTAO calibrations and daytime tomography. The choice was to go with a rectangular phase screen, whose length is dictated by the following factors:
\begin{itemize}
    \item The translation range possible within the space constraints at the required location in the TelSim.
    \item To have an equivalent physical extent such that there is no statistical correlation between both ends of the phase screen which is illuminated by the pupil plane beam. This corresponds to a phase screen length of more than two pupil diameters\cite{correia2014}.
    \item The height of the turbulence on the phase screen, perpendicular to the surface of the AO bench, shall be at least 36.5mm to accommodate the extent of the beam at the Keck pupil mask.
    \item The MK median seeing profile shows a fractional ground-layer contribution of 0.517 with an integrated seeing Fried parameter value of 14.68 cm. With the DM providing the ground-layer turbulence contribution ($<$ 1 km) and the phase screen providing the rest, the phase screen should have a Fried parameter of 22,72 cm.
\end{itemize}
The phase screen manufacturing technique at UCSC involves the strategic application of clear acrylic paint onto an Edmund Optics thermoset CR-39 substrate of size 125 x 100 mm. The primary disadvantage of this method is the non-deterministic nature of the resulting phase screen which can have a considerable RMS error as compared to a theoretically simulated Kolmogorov spectrum that will be difficult to quantify before fabrication and the schedule risk in delivery. Rampy 2012\cite{rampy2012} shows a high adherence to the Kolmogorov spatial spectrum for a physical Fried parameter size of 0.8 mm (which corresponds to the 0.75 mm magnified Fried parameter that we need at the pupil plane). The spatial profile of the phase screen was measured using a Quadrature Polarization Interferometer (QPI) with a spatial resolution of 67 pixels/mm, and is shown in Figure \ref{fig:fig9}a. The analysis of the Fried parameter across the different segments and the spatial spectrum of all the segments are shown in Figure \ref{fig:fig9}b and Figure \ref{fig:fig9}c respectively. Although the statistical analysis made over the 18 available segments show a mean $r_0$ of 0.24 m (close to the target of 0.224m) computed from tilt-removed fit to Noll’s variances, the modal decomposition shows that the phase screen has higher energy in the high frequencies and lower energies in the low frequencies as compared to a Kolmogorov model.

\begin{figure}
    \centering
    \subfigure[]{\includegraphics[width=0.9\textwidth]{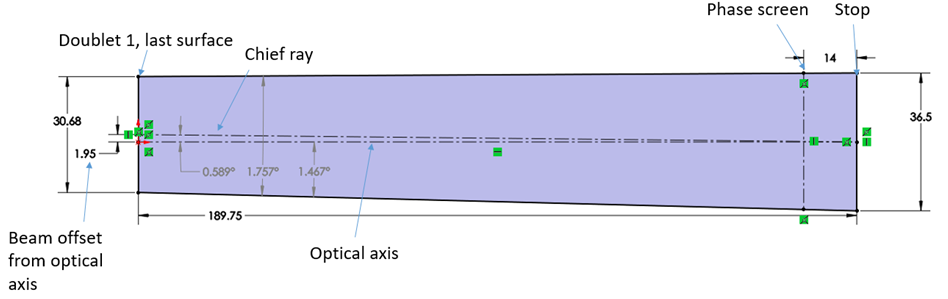}} 
    \subfigure[]{\includegraphics[width=0.43\textwidth]{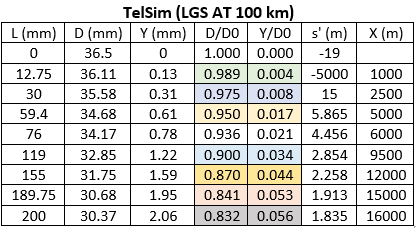}} 
    \subfigure[]{\includegraphics[width=0.3\textwidth]{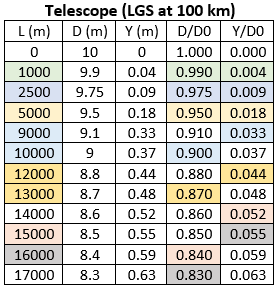}} 
    \caption{(a) Beam path of one of the four LGS beams (located at the LGS focus) between the first achromat and the Keck pupil mask (as modeled with Zemax). The phase screen will be translated along this section of the beam to simulate the presence of turbulence at the different elevations. (b) Beam shift and beam diameter due to focal anisoplanatism when the phase screen is translated across the pupil plane beam path (c) Characteristics of the focal anisoplanatism at the meta-pupils located at the different atmospheric layers.}
    \label{fig:fig10}
\end{figure}

\subsubsection{Focal Anisoplanatism and Tomography}
\label{section:FA}

Figure \ref{fig:fig10}a shows the beam from one of the off-axis LGS fibers from the last surface of the achromat to the 36.5 mm diameter Keck pupil stop. Ground-layer turbulence would be produced if the phase screen were at the at the pupil stop. As the phase screen is moved away from the stop the LGS beam shift on the phase screen corresponds to the shift at higher turbulence altitudes. 

Figure \ref{fig:fig10}b provides the beam shift and beam divergence as a function of the distance between the achromat to the pupil mask, while Figure \ref{fig:fig10}c provides the same parameters for one of the 7.6" off-axis LGS beams ray-traced through the atmosphere. The parameter X in Figure \ref{fig:fig10}b provides the atmospheric layer in Figure \ref{fig:fig10}c for which the optical beam parameters are the closest (D/D0 and Y/Y0). Using a 100 mm translation stage, we will be able to move the phase screen from an equivalent elevation of 5 km to 12 km. The abbreviations in Figure \ref{fig:fig10}b and Figure \ref{fig:fig10}c are as follows:\\
L = Atmospheric layer elevation.\\
D = Beam diameter of each off-axis LGS beam meta-pupil at the specified elevation, L (cone effect makes the D shrink at higher altitudes).\\
Y = Beam shift of the off-axis LGS beam meta-pupil from the optical axis at the specified elevation, L.

\section{Tomographic Algorithm}

\begin{figure}[t]
\includegraphics[width=1\textwidth]{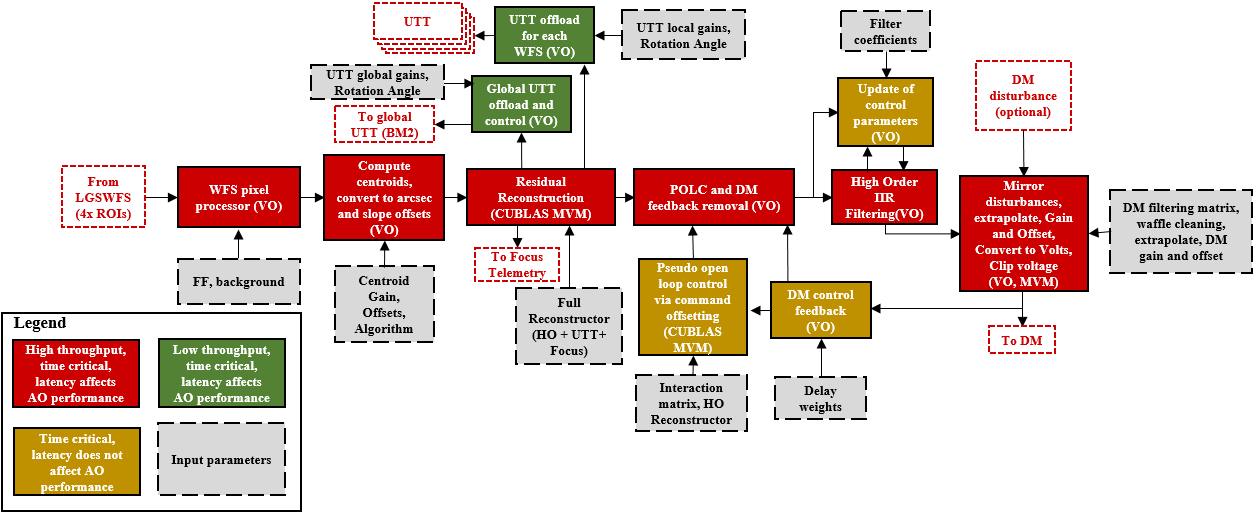}
\centering
\caption{LTAO block diagram with each block corresponding to a GPU kernel operation.}
\label{fig:fig11}
\end{figure}

\begin{figure}
    \centering
    \subfigure[]{\includegraphics[width=0.53\textwidth]{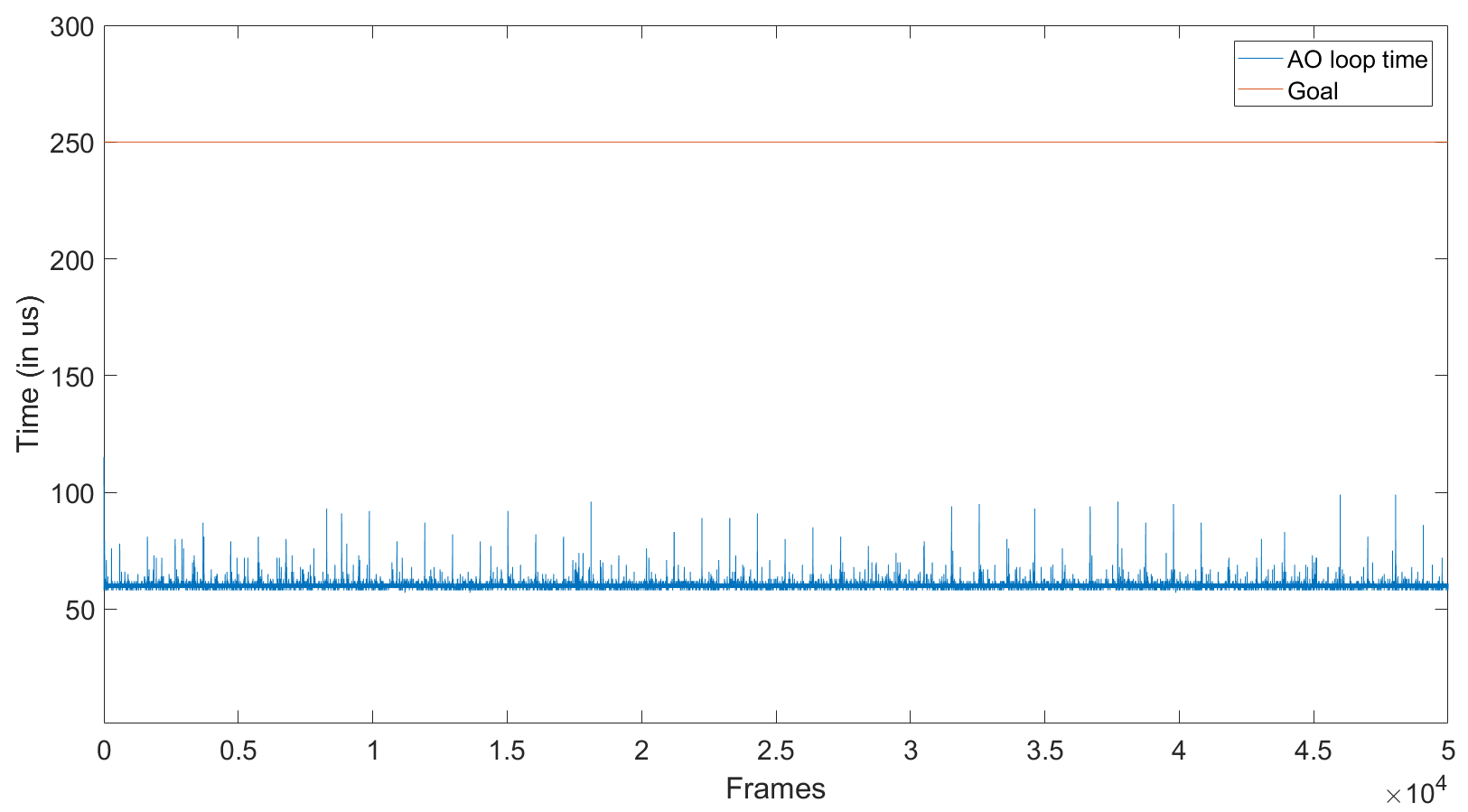}} 
    \subfigure[]{\includegraphics[width=0.43\textwidth]{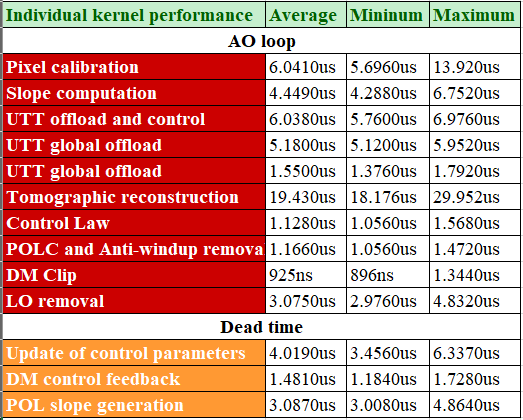}} 
    \caption{(a) Latency of the high throughput time critical stream over 10k frames (b) Latency statistics for the individual blocks in the high-order loop. The latency of the dead time blocks is not included in the RTC timing performance.}
    \label{fig:fig12}
\end{figure}

\label{section:tomo}
The purpose of the daytime calibration infrastructure is to ascertain the performance obtained from the model-based AO reconstruction that is advocated for the LTAO operation. The LTAO algorithms are considerably different from the NGS and single LGS cases in that the advocated regulator is POLC (pseudo open-loop controller) using a MMSE (minimum mean-square error) tomographic reconstructor\cite{Correia15, van2021}. The latter is partially based on a model integrating both the geometric system and the statistical atmospheric parameters. Chin 2022\cite{Chin22} describes the design and preliminary performance results from the NGS and LGS integration of the RTC designed by Microgate in collaboration with Swinburne University and Australian National University. The modifications in the RTC for the LTAO mode (as compared to the single-LGS mode) include the following:
\begin{itemize}
    \item Pixel processing and slope computation of four region-of-interests (ROIs) corresponding to the 4 LGS pupils on the wavefront sensor.
    \item Processing of LGS jitter correction for five uplink tip-tilt (UTT) mirrors (including one for global motion).
    \item An additional POLC step to be operated in the dead-time of the RTC loop, and an MMSE based AO reconstruction.
\end{itemize}
The RTCS software, which will be delivered by Microgate, is broadly divided into the Interface Module (IM) and the Compute Engine (CE). Figure \ref{fig:fig11} shows the CE blocks that will form the LTAO mode of the RTC, with each block corresponding to a single GPU kernel implementation.

A Dell Precision 5820 workstation with an EVGA Geforce GTX 1080TI GPU was procured for the functional testing of the LTAO kernels before they are integrated into Microgate’s RTC. Although the performance of the procured hardware is lower than that of the RTC hardware which is going to be delivered by Microgate, the functionality (primarily based on the CUDA 10.2 architecture) of the two platforms are mostly similar, facilitating a seamless transition of the kernels to host hardware supplied by Microgate. A prototyping code consisting of the complete high order loop implementation (Figure \ref{fig:fig11}) was created to operate on the prototyping GPU using CUDA and associated (CUBLAS) libraries, and was functionally verified with a CPU based implementation of the loop and representative dummy values for the WFS input pixels. The results presented here does not include the latency introduced by hardware interfaces (WFS, UTT and DM), nor does it include updating the soft real-time parameters. The high-order reconstruction uses an MVM of the slopes with a matrix of size 2432 (total number of slopes) by 359 (349 actuators and 10 UTT outputs). Figure \ref{fig:fig12}a shows the total RTC timing latency for each LTAO frame over 10k frames. The total latency will always be more than the sum of the latencies of the individual blocks since the individual block latencies (shown in Figure \ref{fig:fig12}b) does not include kernel launch and memory transfer timings.

\section{Daytime calibration and Testing}
\label{section:daytime}

The current section describes the phasing of hardware integration and the nature of tests which are planned for the validation of LTAO algorithms and to ascertain the AO performance using the telescope and turbulence simulator.

\subsection{Hardware Integration and Testing}
\begin{enumerate}
    \item \textbf{Operationalization of the new light source for AO modes:} Once the new source box is operationalized, this phase will involve the testing and integration of the new sources on the AO bench for supporting NGS, single LGS AO and eventually LTAO science operations with the target of retiring the old source. 
    \item \textbf{Validation of the PRO for daytime testing:} This phase is designed to ensure that the new PRO can be used to support NGS and single LGS AO science operations at first, and LTAO mode after that.
    \item \textbf{Installation of the TelSim and validation for existing AO modes:} This phase involves the verification of the alignment of both the telescope and turbulence simulator on the AO bench, accompanied by calibration for the phase screen altitude.
\end{enumerate}

\subsection{LTAO Functional Verification}
The phasing plan described here is intended to test the two substantially different parts of the tomographic implementation (multiple LGS and POLC) from the existing modes before combining the functionality of both phases into the full tomographic implementation. 
\begin{enumerate}
    \item \textbf{Single LGS POLC:} This phase will focus on testing the single LGS reconstructor in pseudo open loop control mode by injecting the pseudo-open loop slopes into the reconstruction, while only requiring the optical infrastructure used for single LGS.  
    \item \textbf{Averaged correction with multiple LGS:} This phase will focus on our ability to close the high-order AO loop with four LGS sources and to close the LGS jitter loop associated with the five UTTs. This phase makes ground-layer AO correction possible with four lasers without the implementation of tomographic projection. The multi-LGS averaging performance should be better than the single LGS implementation.
    \item \textbf{Tomographic POLC:} The final phase will combine the multi-LGS functionality in step 2 with the POLC reconstruction in step 1 to have the full tomographic pipeline (as shown in Figure \ref{fig:fig11}).
\end{enumerate}

\subsection{Error Breakdown and Performance Estimation}
Once the functional verification and latency benchmarks of the LTAO pipeline are completed, we will use tools to test the optical performance of the pipeline and to help in the testing of the variations of filter parameters and reconstructors for the best performance.

Performance estimates are to be established via two parallel methods:
\begin{itemize}
    \item Computation of DM-produced wavefront variances (in absolute and differential modes wherever required)
    \item Strehl-ratio on the science detector (OSIRIS imager) from which the wavefront error is computed using the inverse Maréchal relationship
\end{itemize}

We are in the early stages of formulating experimental methods for ascertaining the different error budgets, some of which are listed below:
\begin{itemize}
    \item Fitting error
    \begin{itemize}
        \item Closing the loop on a static section of the phase screen, the fitting needs to be scaled by the screen strength at the static location.
        \item Computed from the PSF wings and compared to the commonly accepted analytic expression with adapted scaling constant specific to the Keck DM.
        \item Application of a high-order aberration to the DM but only correct the DM with every 2nd actuator.
    \end{itemize}
    \item Aliasing error
    \begin{itemize}
        \item Compare open-loop reconstructed DM-produced and phase screen produced wavefronts removed from the reconstruction error. 
        \item The wavefronts are computed from the product of the DM commands by the DM influence functions
        \item The turbulence projected on the DM shall have the same turbulence parameters as the as-built phase-screen. Open-loop is justified here since the phase screen turbulence strength is within the  linearity range of the WFS.
    \end{itemize}
    \item Bandwidth Error
    \begin{itemize}
        \item Indirect measurement from Rejection Transfer Function (RTF) and model phase temporal spectrum: RTF computed from the ratio of open-to-closed loop noise
        \item Direct measurement with DM induced translating turbulence (generate no aliasing or fitting error). Reconstruct measurements and compute bandwidth error on a very bright source.
    \end{itemize}
    \item Measurement Error: Fit noise as an expression of magnitude and spot sizes, with the choice of having multiple fiber sizes.
    \item Tomographic Error: With the AS fibers at finite altitudes, we can directly estimate the combination of focal and angular anisoplanatism (the full tomographic error). 
    \begin{itemize}
        \item Close the AO loop on a bright source with the phase screen in the optical path without any turbulence injected on the DM. 
        \item Drive the DM in the full tomographic reconstruction and in multi-LGS averaging modes.
        \item Allow enough time for the system to settle before telemetry acquisition (making it quasi-static to remove any bandwidth errors)
        \item Translate the phase-screen to a new position (1/8th of a subaperture) and repeat all steps for the phase-screen conjugated at different altitudes.
    \end{itemize}
The combined performance should estimate should be the best at NGS AO, followed by full tomography (with MMSE and POLC), followed by an averaging reconstructor over the four LGS. The performance outputs of all these modes will be compared to the predicted model. 

\end{itemize}

\section{Current Status}
\label{section:status}

\begin{figure}
    \centering
    \subfigure[]{\includegraphics[width=0.53\textwidth]{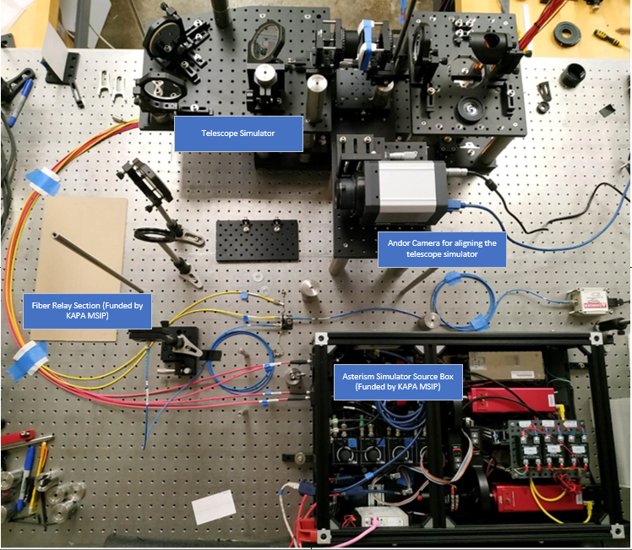}} 
    \subfigure[]{\includegraphics[width=0.41\textwidth]{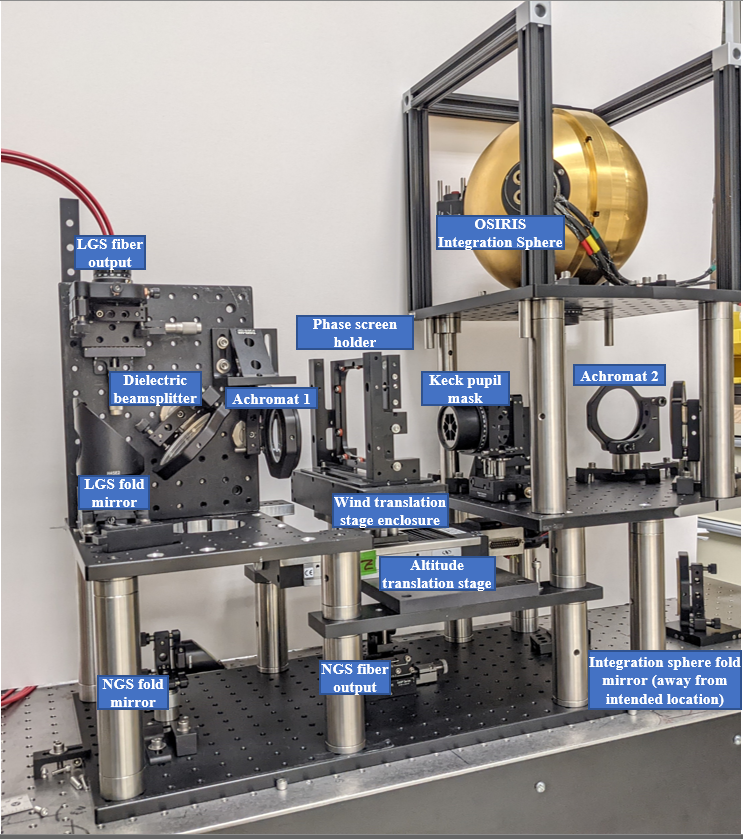}}
    \subfigure[]{\includegraphics[width=0.2\textwidth]{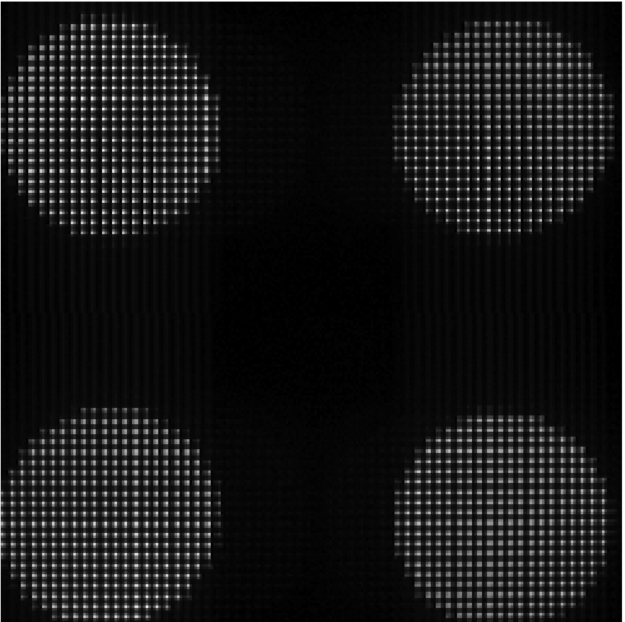}} 
    \subfigure[]{\includegraphics[width=0.7\textwidth]{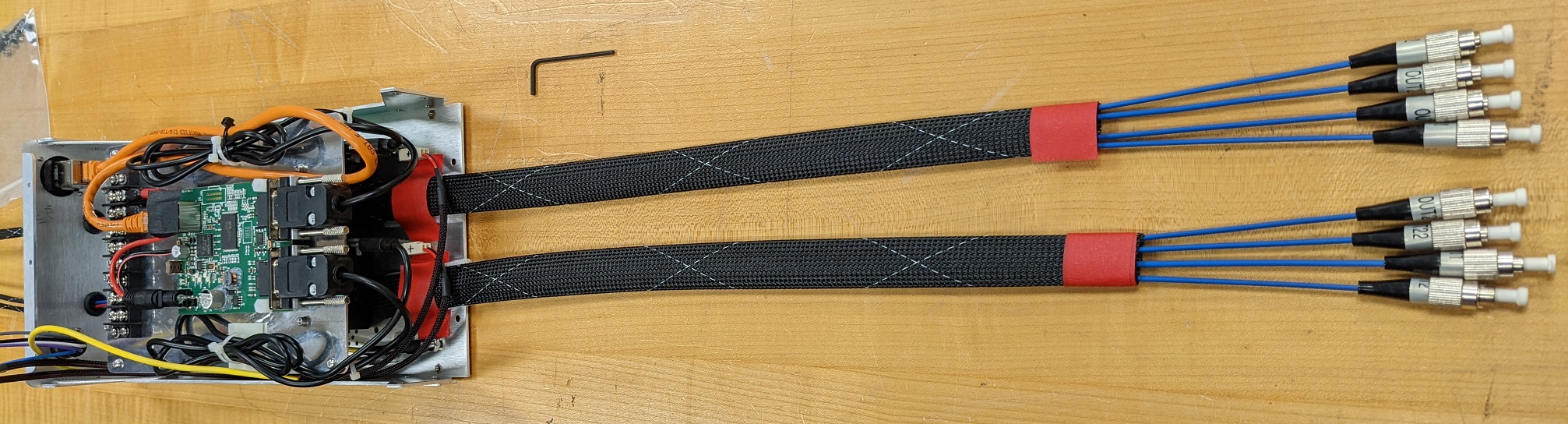}}
    \caption{(a) The AS fiber holder asterism radius test setup in the lab with the source box providing the light sources and the TelSim reimaging the asterism on to an Andor Marana sCMOS camera (b) TelSim Assembly in the lab with labeled parts. (c) Image of the four LGS pupils on the OCAM2K wavefront sensor as relayed by the PRO in the lab. (d) Agiltron fiber splitter and attenuator being assembled in the lab, with the eight LGS fiber outputs shown.}
    \label{fig:fig13}
\end{figure}

\begin{itemize}
    \item The detailed design review (DDR) pertaining to the daytime calibration infrastructure was completed at the end of 2020, but there were substantial design changes to the TelSim to accommodate the placement of the LGS sources at the finite altitude. The original TelSim design only accounted for angular anisoplanatism for the LGS sources as compared to the NGS.
    \item The AS source box and fiber relay have been assembled (Figure \ref{fig:fig13}a), tested and installed with the Keck 1 AO bench. Operational transition and retirement of the old light source is in process. The Agiltron photonic fiber splitter and attenuator (Figure \ref{fig:fig13}d shows the eight output fibers coming out from the two Agiltron devices) has been functionally tested and is in the process of installation and interconnection to the summit systems.
    \item Plate scale and pupil separation tests for the PRO have been completed in the lab. PRO installation will follow after the operationalization of the NGS and LGS modes of the RTC. Figure \ref{fig:fig13}c shows the four pupils (slightly overfilled) on the OCAM2K wavefront sensor.
    \item The TelSim is undergoing assembly and alignment in the lab (Figure \ref{fig:fig13}b).
    \item The tomographic algorithm is being integrated into the RTC in collaboration with Microgate, Swinburne University and Australian National University.
\end{itemize}

\acknowledgments 
The W. M. Keck Observatory is operated as a scientific partnership among the California Institute of Technology, the University of California, and the National Aeronautics and Space Administration. The Observatory was made possible by the generous financial support of the W. M. Keck Foundation. The authors wish to recognize and acknowledge the very significant cultural role and reverence that the summit of Maunakea has always had within the indigenous Hawaiian community. We are most fortunate to have the opportunity to conduct observations from this mountain. The Keck II realtime controller is funded by the NSF Major Research for Instrumentation Program award AST-1727071 (PI: Wizinowich). Funding support for the KAPA system is provided by the National Science Foundation Mid-Scale Innovations Program award AST-1836016 (PI: Wizinowich). We would like to thank the KAPA science team of Andrea Ghez, Tuan Do, Mark Morris, Shelley Wright, Tucker Jones, Claire Max, Michael Liu and Dimitri Mawet for their support.

\bibliography{main} 
\bibliographystyle{spiebib} 

\end{document}